# Conjugate gradient methods in micromagnetics

J. Fischbacher,[1] Alexander Kovacs,[1] Harald Oezelt,[1] T. Schrefl,[1,a] L. Exl,[2,3] J. Fidler[3], D. Suess[4], N. Sakuma,[5,6] M. Yano,[5,6] A. Kato,[5,6] T. Shoji[5,6], and A. Manabe[6]

[1]*Center for Integrated Sensor Systems, Danube University Krems, 2700 Wiener Neustadt, Austria*

[2]*Faculty of Mathematics, University of Vienna, Vienna, 1090 Wien, Austria*

[3]*Institute for Solid State Physics, TU Wien, 1040 Wien, Austria*

[4]*CD-Laboratory for Advanced Magnetic Sensing and Materials, TU Wien, 1040 Vienna, Austria*

[5]*Toyota Motor Corporation, 1200 Mishuku, Susono, Shizuoka 410-1193, Japan*

[6]*Technology Research Association of Magnetic Materials for High-efficiency Motors (Mag-HEM) Higashifuji-Branch, 1200 Mishuku, Susono, Shizuoka 410-1193, Japan*

Conjugate gradient methods for energy minimization in micromagnetics are compared. When the step length in the line search is controlled, conjugate gradient techniques are a fast and reliable way to compute the hysteresis properties of permanent magnets. The method is applied to investigate demagnetizing effects in $NdFe_{12}$ based permanent magnets. The reduction of the coercive field by demagnetizing effects is $\mu_0 \Delta H = 1.4$ T at 450 K.

**I. INTRODUCTION**

The computation of hysteresis properties of large ferromagnetic systems such as sensor elements or permanent magnets require fast and reliable solvers. Hysteresis simulations are based on the theory of micromagnetics "Brown (1963)". The primary purpose of these simulations is to understand the influence of the microstructure on magnetization reversal. In this work we are focusing on the role of demagnetizing fields in platelet shaped grains of permanent magnets. We also describe the key elements of a micromagnetic solver suitable for simulating large magnetic systems.

After discretization of the total Gibbs free energy with finite elements or finite differences the states along the demagnetization curve can be computed by subsequent minimization of the energy for decreasing applied field as outlined in "Kinderlehrer (1997)". The system is in a metastable state. A change of the applied field shifts the position of the local energy minimum. At a critical field, the magnetization becomes unstable. An irreversible switching occurs which is seen as a kink in the demagnetization curve. Then the system either accesses a different metastable state or if fully reversed the magnetization is in a stable state. A reliable numerical method for energy minimization must track all local minima along the demagnetization curve. The resulting algebraic minimization problem is large. Typically the number of unknowns is in the order of 10 to 50 million for a model magnet consisting of around 10 grains. Therefore fast numerical methods are required to obtain results in a reasonable time. In finite element micromagnetics the conjugate gradient method "Koehler and Fredkin (1992)" has been used to compute the hysteresis curve of magnets. These techniques go back to the seminal paper of "Cohen and co-workers (1989)" on relaxation and gradient method for the calculation of the molecular orientation in liquid crystals. There are two reasons why it is useful to revisit conjugate gradient methods in micromagnetics:

(1) Since the early use of the conjugate gradient methods in the early 1990s in micromagnetics several new conjugate gradient schemes have been developed that show better properties than the original methods.

(2) The advance of hardware makes it possible to apply numerical micromagnetics for large systems. This requires algorithms that correctly track the local minima of the system.

---

[a] Electronic mail: tschrefl@gmail.com

"Exl et al. (2014)" and "Furuya et al. (2015)" introduced steepest descent methods for large scale micromagnetics. In this work we focus on conjugate gradient methods and required modifications for their use in numerical micromagnetic simulations. We test the method by simulating magnetization reversal in a permanent magnet and in a soft magnetic permalloy element. The grains of the permanent magnet are assumed to be separated by a non-magnetic grain boundary phase. Switching off the demagnetizing field in the micromagnetic simulations the permanent magnet turns into a system of Stoner-Wohlfarth particles. Thus we can test the results obtained for different conjugate gradient methods against an analytic solution. Comparing the numerical results which were computed with and without magnetostatics we can quantify the reduction of coercivity owing to the self demagnetizing field. In soft magnetic thin film elements metastable magnetic states are found along the demagnetization curve. These states may give rise to jitter in magnetic sensor elements. We will show how conjugate gradient methods need to be tuned, in order to access these states during the computation of the demagnetization curve. For comparison we also include results obtained with a limited memory quasi-Newton method into our study.

## II. NUMERICAL BACKGROUND

### A. Discrete micromagnetics

Let us introduce the following notation. Each node of a finite element mesh or cell of a finite difference scheme holds a unit magnetization vector **m**. We gather these vectors into the vector **x** which has the dimension $3n$, where $n$ is the number of nodes or cells so that $x_{3i} = m_{i,x}$, $x_{3i+1} = m_{i,y}$, and $x_{3i+2} = m_{i,z}$. The node or cell index $i$ runs from 0 to $n-1$. Similarly the external field and the demagnetizing fields at the nodes or cells are $\mathbf{h}_{\text{ext}}$ and $\mathbf{h}_d$, respectively. Then the total Gibbs free energy is

$$F(\mathbf{x}) = \frac{1}{2}\mathbf{x}^T \mathbf{C}\mathbf{x} - \frac{1}{2}\mathbf{h}_d^T \mathcal{M}\mathbf{x} - \mathbf{h}_{ext}^T \mathcal{M}\mathbf{x} \qquad (1)$$

The first, second, and last term of the right hand side of (1) are the sum of the exchange and anisotropy energy, the magnetostatic self energy, and the Zeeman energy, respectively. Within the framework of the finite element method with linear basis functions $\varphi_i$ the matrix entries in $\mathbf{C}$ can be derived as follows. Approximated on the finite element grid, the unit magnetization vector is $\mathbf{m}(\mathbf{r}) = \sum_{i=0}^{n-1} \mathbf{m}_i \varphi_i(\mathbf{r}) = \mathbf{m}_i \varphi_i(\mathbf{r})$. It's Cartesian components are $m_k(\mathbf{r}) = x_{3i+k} \varphi_i(\mathbf{r})$ with $k = 0, 1, 2$. Here we use the Einstein summation convention and the notation $\varphi_{i,k} = \partial \varphi_i / \partial r_k$ for the partial derivative of $\varphi_i$ with respect to direction $k$. The sum of the exchange and anisotropy energy of a ferromagnetic body is

$$F_{exani}(\mathbf{m}) = \int_V A\{(\nabla m_0)^2 + (\nabla m_1)^2 + (\nabla m_2)^2\} dV - \int_V K(u_k m_k)^2 dV \qquad (2)$$

where $A$ and $K$ are the exchange constant and the uniaxial magneto-crystalline anisotropy constants. $\mathbf{u} = (u_0, u_1, u_2)^T$ is the unit vector along the anisotropy direction. Replacing $m_k$ by $x_{3i+k}\varphi_i(\mathbf{r})$ we obtain

$$F_{exani}(\mathbf{m}) = \int_V A\{x_{3i}\varphi_{i,s}\varphi_{j,s}x_{3j} + x_{3i+1}\varphi_{i,s}\varphi_{j,s}x_{3j+1} + x_{3i+2}\varphi_{i,s}\varphi_{j,s}x_{3j+2}\} dV$$
$$- \int_V K(x_{3i+k}u_k\varphi_i\varphi_j u_l x_{3j+l}) dV \qquad (3)$$

the indices $i, j$ run from 0 to $n-1$, and $s, k, l$ run from 0 to 2. Comparing (1) and (3) we get the elements of the matrix $\mathbf{C}$

$$C_{3i+k,3j+l} = \int_V 2A\delta_{kl}\varphi_{i,s}\varphi_{j,s} dV + \int_V 2K(u_k\varphi_i\varphi_j u_l) dV. \qquad (4)$$

The matrix $\mathcal{M}$ is a diagonal matrix whose entries are the modulus of the magnetic moment associated with the node or cell $i$. In a finite difference setting we define

$$\mathcal{M}_{3i,3i} = \mathcal{M}_{3i+1,3i+1} = \mathcal{M}_{3i+2,3i+2} = \int_{V_i} \mu_0 M_s dV. \qquad (5)$$

$M_s$ is the magnetization and the integral is over the volume of cell $i$. For a finite element discretization with linear basis functions $\varphi_i$ the $\mathcal{M}_{3i+k,3i+k}$ are the solution of the linear system

$$\int_V \mathcal{M}_{3i+k,3i+k}\varphi_i dV = \int_V \mu_0 M_s \varphi_i dV. \quad k = 0, 1, 2 \tag{6}$$

The matrices **C** and $\mathcal{M}$ depend on the intrinsic magnetic properties, the geometry, and the computational grid. Similar expression for the discretized micromagnetic energy are given by "Koehler and Fredkin (1992)" and "Schrefl and co-workers (2007)". A basic assumption of micromagnetics is the length of the magnetization vector is a function of temperature only and does not depend on the applied field, $|\mathbf{M}| = M_s(T)$ "Brown (1963)". This translates into the following contraints

$$|\mathbf{m}_i| = 1. \qquad i = 0, \ldots, n-1 \tag{7}$$

The minimization of (1) with respect to **x** subject to (7) gives a metastable or stable equilibrium state of the magnetization.

**B. Conjugate gradient methods**

The conjugate gradient method is an efficient method for large scale optimization. The conjugate gradient method was originally introduced by "Hestenes and Stiefel (1952)" for the solution of a linear system of equation $\mathbf{Cx} = -\mathbf{b}$, which is equivalent to minimize the quadratic function $Q(\mathbf{x}) = (1/2)\mathbf{x}^T\mathbf{Cx} + \mathbf{b}^T\mathbf{x} + \mathbf{c}$. The conjugate gradient method has been extended for the minimization of non-quadratic functions by "Fletcher and Reeves (1965)", "Polyak (1967)",.and "Polak and Ribiere (1969)". The conjugate gradient method is given in algorithm 1.

**Algorithm 1: Nonlinear conjugate gradient method:**

Minimize $F(\mathbf{x})$:

take an initial guess $\mathbf{x}_0$ and compute the energy gradient $\mathbf{g}_0 = \nabla F(\mathbf{x}_0)$ (8)

set the initial search direction $\mathbf{d}_0 = -\mathbf{g}_0$ (9)

for $i = 0, 1, 2, 3 \ldots$ do

    compute a step length $\alpha_i$ by minimizing $F(\mathbf{x}_i + \alpha_i \mathbf{d}_i)$ with respect to $\alpha_i$ (line search)

    set new solution value $\quad \mathbf{x}_{i+1} = \mathbf{x}_i + \alpha_i \mathbf{d}_i$ (10)

    compute energy gradient $\quad \mathbf{g}_{i+1} = \nabla F(\mathbf{x}_{i+1})$ (11)

    exit if convergence criteria are fulfilled

    compute a new search direction $\quad \mathbf{d}_{i+1} = -\mathbf{g}_{i+1} + \beta_i \mathbf{d}_i$ (12)

There are various different ways to compute the factor $\beta_i$ in (12). Most prominent are the Fletcher-Reeves conjugate gradient method, which uses

$$\beta_i^{\text{FR}} = \frac{\mathbf{g}_{i+1}^T \mathbf{g}_{i+1}}{\mathbf{g}_i^T \mathbf{g}_i}, \tag{13}$$

the Polak-Ribiere-Polyak conjugate gradient method

$$\beta_i^{\text{PRP}} = \frac{(\mathbf{g}_{i+1} - \mathbf{g}_i)^T \mathbf{g}_{i+1}}{\mathbf{g}_i^T \mathbf{g}_i}, \tag{14}$$

and the Hestenes-Stiefel conjugate gradient method

$$\beta_i^{\text{HS}} = \frac{(\mathbf{g}_{i+1} - \mathbf{g}_i)^T \mathbf{g}_{i+1}}{(\mathbf{g}_{i+1} - \mathbf{g}_i)^T \mathbf{d}_i}. \tag{15}$$

The search direction of the Hestenes-Stiefel fulfills

$$(\mathbf{g}_{i+1} - \mathbf{g}_i)^T \mathbf{d}_{i+1} = 0, \tag{16}$$

which can be seen by multiplying (12) with $(\mathbf{g}_{i+1} - \mathbf{g}_i)$ and using (15). Equation (16) is called the conjugacy condition. In the quadratic case $(\mathbf{g}_{i+1} - \mathbf{g}_i)$ is parallel to $\mathbf{C}\mathbf{d}_i$ and thus (16) reads $\mathbf{d}_i^T \mathbf{C} \mathbf{d}_{i+1} = 0$ which means the search direction are conjugate.

Since the invention of the conjugate gradient many modified conjugate gradient methods have been proposed "Dai (2011)". Some of them show very good properties for the solution of large optimization problems. "Andrei (2007)" shows that their performance is comparable with fast quasi-Newton solvers. In the remainder of the paper we will test different conjugate gradient methods for their use in micromagnetics.

From the chain rule we have $\partial F(\mathbf{x}_i + \alpha \mathbf{d}_i)/\partial \alpha|_{\alpha=0} = \mathbf{g}_i^T \mathbf{d}_i$ and $\partial F(\mathbf{x}_i + \alpha \mathbf{d}_i)/\partial \alpha|_{\alpha=\alpha_i} = \mathbf{g}_{i+1}^T \mathbf{d}_i$. For an minimization method we require the search direction (12) to be a descent direction: $\mathbf{g}_{i+1}^T \mathbf{d}_{i+1} < 0$. When the line search is exact $\mathbf{g}_{i+1}^T \mathbf{d}_i = 0$. Then it follows from multiplying (12) with $\mathbf{g}_{i+1}^T$ that $\mathbf{g}_{i+1}^T \mathbf{d}_{i+1} = -|\mathbf{g}_{i+1}|^2$. An exact line search is normally to expensive as it requires many function and/or gradient evaluations. Nonlinear conjugate gradient methods normally apply an inexact line search that satisfies the strong Wolfe conditions:

$$F(\mathbf{x}_i + \alpha_i \mathbf{d}_i) \leq F(\mathbf{x}_i) + c_1 \alpha_i \mathbf{g}_i^T \mathbf{d}_i \tag{17}$$

$$|\mathbf{g}_{i+1}^T \mathbf{d}_i| \leq c_2 |\mathbf{g}_i^T \mathbf{d}_i| \tag{18}$$

With $0 < c_1 < c_2 < \frac{1}{2}$ "Nocedal and Write (2006)". The step size $\alpha_i$ is chosen such that the new point has a sufficiently smaller energy and sufficiently smaller gradient along the search direction than the previous point. In particular, condition (18) guarantees that the step size is large enough and that the new point is close to a local minimum. A line search algorithm that finds a step length which fulfills the strong Wolfe conditions is given by "Moré and Thuente (1994)". Alternatively, a modified Armijo line search is often used in conjugate gradient methods "Bartholomew-Biggs (2008)".

A well known modification of the Polak-Ribiere-Polyak conjugate gradient method is to set $\beta$ to zero if $\beta^{\mathrm{PRP}} < 0$:

$$\beta^{\mathrm{PRP+}} = \max(\beta^{\mathrm{PRP}}, 0). \tag{19}$$

This ensures that (12) is a descent direction with a line search that satisfies modified Wolfe conditions. One class of modified conjugate gradient method changes (12) so that the new search directions are sufficient descent independent of the line search. One such method was proposed by "Cheng (2007)"

$$\mathbf{d}_{i+1} = -\left(1 + \beta_i^{\mathrm{PRP+}} \frac{\mathbf{g}_{i+1}^{\mathrm{T}} \mathbf{d}_i}{|\mathbf{g}_{i+1}|^2}\right) \mathbf{g}_{i+1} + \beta_i^{\mathrm{PRP+}} \mathbf{d}_i. \tag{20}$$

The search direction (18) always fulfills $\mathbf{g}_{i+1}^{\mathrm{T}} \mathbf{d}_{i+1} = -|\mathbf{g}_{i+1}|^2$. When the line search is exact the second term in the bracket of the right hand side vanishes. A second class of modified conjugate gradient methods changes (12) so that the conjugacy condition (16) is always satisfied independent of the line search "Lukšan and co-workers (2008)". A modified Polak-Ribiere-Polyak method with guaranteed conjugacy is

$$\mathbf{d}_{i+1} = \begin{cases} -\frac{(\mathbf{g}_{i+1}-\mathbf{g}_i)^{\mathrm{T}} \mathbf{d}_i}{|\mathbf{g}_i|^2} \mathbf{g}_{i+1} + \beta_i^{\mathrm{PRP}} \mathbf{d}_i & \text{if } \beta_i^{\mathrm{PRP}} > 0 \\ -\mathbf{g}_{i+1} & \text{otherwise} \end{cases}. \tag{21}$$

We also test a modified Hestenes-Stiefel conjugate gradient method

$$\beta_i^{\mathrm{ZA}} = \begin{cases} \beta_i^{\mathrm{HS}} & \text{if } |\mathbf{g}_{i+1}|^2 > |\mathbf{g}_{i+1}^{\mathrm{T}} \mathbf{g}_i| \\ 0 & \text{otherwise} \end{cases} \tag{22}$$

which was proposed by "Salleh and Alhawarat (2016)". Hybrid conjugate gradient method combine two or more methods. "Touati-Ahmed and Storey (1990)" suggested

$$\beta_i^{\text{TaS}} = \begin{cases} \beta_i^{\text{PRP}} & \text{if } 0 \leq \beta_i^{\text{PRP}} \leq \beta_i^{\text{FR}} \\ \beta_i^{\text{FR}} & \text{otherwise} \end{cases}, \qquad (23)$$

"Hu and Storey (1991)" proposed

$$\beta_i^{\text{HuS}} = \max(0, \min(\beta_i^{\text{PRP}}, \beta_i^{\text{FR}})). \qquad (24)$$

Setting $\beta = 0$, such as it is done occasionally in the PRP+ , ZA, and HuS method, restarts the conjugate gradient algorithms with $\mathbf{d}_{i+1} = -\mathbf{g}_{i+1}$. The convergence of the conjugate gradient method can be improved by periodic restarts. If the problem size is large we can restart the algorithm after every *m* steps.

## C. Quasi Newton methods

In a quasi-Newton method the search direction is computed using an approximation of the inverse of the Hessian matrix. In particular equation (12) is replaced by

$$\mathbf{d}_{i+1} = -\mathbf{B}_{i+1}^{-1} \mathbf{g}_{i+1}, \qquad (25)$$

whereby the Hessian $\mathbf{B}_{i+1}$ is updated during the iterations. In practical applications of quasi-Newton methods instead of the Hessian matrix $\mathbf{B}$ an approximation of the inverse is stored. "Liu and Nocedal (1989)" give an efficient implementation of the quasi-Newton method which stores an approximation of $\mathbf{B}^{-1}$ implicitly using a few vector pairs $(\mathbf{x}_{i+1} - \mathbf{x}_i)$ and $(\mathbf{g}_{i+1} - \mathbf{g}_i)$. Then product $\mathbf{B}_{i+1}^{-1} \mathbf{g}_{i+1}$ can be computed performing a sequence of inner products and vector summations. Usually only about five of the latest vector pairs are used to compute the search direction.

For optimal performance the approximation of the matrices $\mathbf{B}_{i+1}$ should be symmetric and positive definite "Nocedal and Wrigth (2006)". This can be enforced using a line search that fulfills the Wolfe conditions (17) and (18). Some implementations are less strict on the line search but skip the update of the inverse of the Hessian by $\mathbf{B}_{i+1}^{-1} = \mathbf{B}_i^{-1}$ when a new vector pair would cause the Hessian to be indefinite. This happens when the curvature condition

$$(\mathbf{g}_{i+1} - \mathbf{g}_i)^{\text{T}} (\mathbf{x}_{i+1} - \mathbf{x}_i) > 0 \qquad (26)$$

is violated. If the updates are skipped to often the curvature information gets poor and the algorithm may fail.

**D. The conjugate gradient method for micromagnetics**

The energy (1) is a quadratic function. However the constraints (7) makes the micromagnetic problem highly nonlinear. "Cohen et al. (1989)" introduced an efficient way to treat the constraints |$\mathbf{m}_i$| = 1 in combination with the conjugate gradient method. The vectors $\mathbf{m}_i$ should stay on the unit sphere. During updates the vectors should only change their direction. Therefore "Cohen et al. (1989)" project the energy gradient on its component perpendicular to the local orientation vector. After each update the orientation vector is normalized. "Exl (2014)" showed that this approach is equivalent to the approach adopted by "Koehler and Fredkin (1992)" who formally replaced $\mathbf{m}$ with $\mathbf{m}/|\mathbf{m}|$ in (1). "Cohen et al. (1989)" used the Polak-Ribiere-Polyak conjugate gradient method. "Koehler and Fredkin (1992)" used the Fletcher-Reeves conjugate gradient method. In micromagnetics the evaluation of energy and its gradient is expensive. "Koehler and Fredkin (1992)" pointed out that too many function and gradient evaluations are prohibitive in micromagnetics. Therefore they propose an inexact line search based on cubic interpolation.

"Cohen et al. (1989)" used the Polak-Ribiere-Polyak conjugate gradient method with periodic restarts for the calculation of the molecular orientation in liquid crystals. "Koehler and Fredkin (1992)" applied the Fletcher-Reeves conjugate gradient methods. They restart the scheme whenever a sudden drop of the total energy occurs. "Koehler and Fredkin (1992)" also point out that a strict convergence criterion is required. They stop the search for the next local minimum when the relative change in the energy between two steps is less than $10^{-10}$.

In our simulation we apply Cohen's approach for treating the constraints (7). Equations (8) and (11) are modified. The gradient is replaced by the projection of $\nabla F$ onto its component perpendicular to $\mathbf{m}$. Then elements of the vector $\mathbf{g}$ are defined as

$$g_{3i+k} = (\nabla F)_{3i+k} - \left[\delta_{ij}(\nabla F)_{3j+l} x_{3j+l}\right] x_{3i+k}. \tag{27}$$

As previously the index $i$ runs over all nodes from 0 to $n$–1. Indices $k$, $l$ denote the Cartesian components and run from 0 to 2. The bracket in (27) is the inner product between the $\nabla F$ and $\mathbf{m}$ at node $i$. We compute the step length by minimizing $F\left(\frac{\mathbf{x}_i + \alpha_i \mathbf{d}_i}{|\mathbf{x}_i + \alpha_i \mathbf{d}_i|}\right)$ with respect to $\alpha_i$ and replace (10) with $\mathbf{x}_{i+1} = \frac{\mathbf{x}_i + \alpha_i \mathbf{d}_i}{|\mathbf{x}_i + \alpha_i \mathbf{d}_i|}$.

With advance in computer hardware micromagnetic simulations of complex granular structures are possible. Such systems may have several local minima that are accessed during magnetization reversal. A simple example is a permanent magnet consisting of multiple isolated grains with slightly different anisotropy directions. Each grain has its distinct switching field leading to kinks in the demagnetization curve. The naïve use of standard library routines for energy minimization may lead to a step length which is too large. Instead of accessing the next local minimum along the demagnetization curve the system will jump to another local minimum. This leads to the switching of two or more grains at one field point. Kinks in the demagnetization curves are missed (see figure 2). Given a magnetic state during at a conjugate gradient iteration we aim to find a step length that is short enough so that no local minima along the demagnetization curve is missed. On the other hand the step should be selected in a way that the iterative minimization scheme makes sufficient progress. We compute the curvature of the energy along the search direction by forward finite differences from the energy gradients and apply a single Newton step in one dimension to get an estimate for the required step size. A second estimate can be found by quadratic extrapolation from the previous function values and the first derivative. We use the minimum of both estimates as the initial step in a backtracking line search that only reduces the step size. Standard line search routines may also increase the line search in order to fulfill the curvature condition (18). Algorithm 2 outlines our line search method.

**Algorithm 2: Backtracking line search with extrapolated initial step:**

compute $\alpha_i$ so that the first Wolfe condition (17) is fulfilled:

compute finite difference interval

$$h = \min\left(\frac{2\sqrt{\epsilon_m}(1+|\mathbf{x}|)}{|\mathbf{d}|}, \frac{\sqrt{\epsilon_m}}{|\mathbf{d}|_\infty}\right) \quad (28)$$

compute initial step length:

$$\alpha_{i,0} = \min\left(\left|\frac{\mathbf{g}_i^T\mathbf{d}_i}{(\mathbf{g}(\mathbf{x}_i+h\mathbf{d}_i)^T\mathbf{d}_i - \mathbf{g}(\mathbf{x}_i)^T\mathbf{d}_i)/h}\right|, \frac{2(F(\mathbf{x}_i)-F(\mathbf{x}_{i-1}))}{\mathbf{g}_i^T\mathbf{d}_i}\right) \quad (29)$$

for $k = 0, 1, 2, 3 \dots$ do

    if $F(\mathbf{x}_i + \alpha_{i,k}\mathbf{d}_i) \leq F(\mathbf{x}_i) + c_1\alpha_{i,k}\mathbf{g}_i^T\mathbf{d}_i$ then step length found: $\alpha_i = \alpha_{i,k}$     (30)

    otherwise reduce step length    $\alpha_{i,k+1} = 0.5\alpha_{i,k}$     (31)

In order to balance the discretization error and the rounding errors the finite difference interval, $h$, is chosen as suggested by "Derreumaux et al. (1994)". In (28) $\epsilon_m$ is the machine precision. We set $c_1 = 0.1$.

For stopping the iterations we apply the convergence criteria proposed by "Gill et al. (1981)". These criteria are based on the function tolerance, $\tau_F$, which is related to the number of correct significant figures, $a$, in the energy. In our implementation the energy is scaled so that $|F| \approx 1$. Following "Koehler and Fredkin (1992)" we set $a = 10$ for most of our computer experiments. We summarize the check for convergence in Algorithm 3. $|\mathbf{x}|_\infty$ is the infinity norm of the vector $\mathbf{x}$.

**Algorithm 3: Check convergence:**

choose the number of required correct significant digits $a$

set the function tolerance
$$\tau_F = 10^{-a} \tag{32}$$

convergence is reached if all of the following three conditions are satisfied:

change in function values small enough $\quad F(\mathbf{x}_i) - F(\mathbf{x}_{i+1}) < \tau_F(1 + |F(\mathbf{x}_{i+1})|) \tag{33}$

change in solution vector small enough $\quad |\mathbf{x}_i \mathbf{x}_{i+1}|_\infty < \sqrt{\tau_F}(1 + |\mathbf{x}_{i+1}|_\infty) \tag{34}$

gradient small enough $\quad |\mathbf{g}_{i+1}|_\infty < \sqrt[3]{\tau_F}(1 + |F(\mathbf{x}_{i+1})|) \tag{35}$

In micromagnetics the torque or the right hand side of the Landau-Lifshitz-Gilbert equation is often used as a stopping criterion. With the above convergence criteria with 10 significant digits the algorithms the normalized torque, we achieve

$$\frac{|\mu_0 \mathbf{M} \times \mathbf{H}_{\text{eff}}|_\infty}{\mu_0 M_s^2} < C \tag{36}$$

at the computed stationary states at all nodes of the finite element mesh. Here $\mathbf{M}$ and $\mathbf{H}_{\text{eff}}$ are the magnetization vector and the effective magnetic field at the nodes of the magnetization vector. The value of the threshold depends on the nature of the sample. We found $C = 0.002$ for permanent magnets and $C = 0.001$ for soft magnetic thin films when $\tau_F = 10^{-10}$.

## III. RESULTS AND DISCUSSION

### E. Permanent magnets

We first compute the demagnetization curve of an assembly of 9 hard magnetic grains. We choose a NdFe$_{12}$ based permanent magnet material which has been investigated by "Suzuki et al. (2016)". In NdFe$_{12}$ the fraction of rare earth is lower than in Nd$_2$Fe$_{14}$B whereas similar or better magnetic properties are expected especially at elevated temperature. In particular at elevated temperatures the magnetization remains higher than in Nd$_2$Fe$_{14}$B. This may help to obtain a large energy density product at the operating temperature of a motor but strong demagnetizing effects may harm the coercive field.

In our sample (see figure 1) the anisotropy axes of the grains are slightly misaligned with an average misalignment angle of 3.46 degrees. The flat sample shape is take on purpose, in order to maximize the effect of demagnetizing fields. We used the following intrinsic magnetic properties for a temperature $T = 450$ K: magnetocrystalline-anisotropy constant $K = 2.37$ MJ/m³, saturation magnetization $\mu_0 M_s = 1.61$ T, exchange constant $A = 9.2$ pJ/m. The sample is discretized with tetrahedral finite elements. The grain boundary phase is assumed to be non-magnetic with a thickness of 5 nm. The grain structure and an enlarged section showing a slice through the finite element is shown in figure. 1. For computing the demagnetizing field we extend the finite element mesh outside the magnetic structure. We use a standard finite element scheme to compute the magnetic scalar potential from which we derive the demagnetizing field. The linear system of equation for the magnetic scalar potential is solved with an algebraic multigrid method.

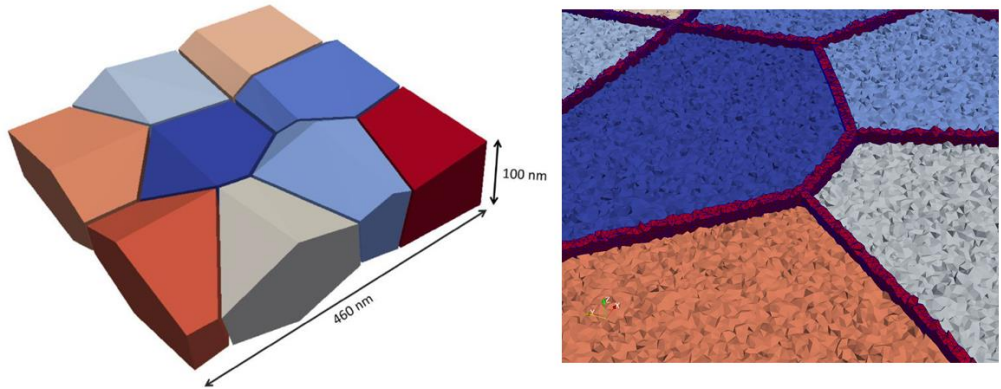

FIG. 1. Granular sample for the computation of the demagnetization curve. The sample size is 460 x 460 x 100 nm³. It consist of 9 grains which are separated by a non-magnetic grain boundary phase as shown on the right hand side. The thickness of the grain boundary phase is 5 nm.

Before computing demagnetizing effects in this sample we want to check our micromagnetic solver and the influence of different energy minimization schemes on the results. In particular we can choose between different energy minimization methods (quasi Newton or various variants of the conjugate gradient method) and different line search methods. Table I lists the various methods used.

TABLE I. Energy minimization schemes within this work. The columns list the following information. **method**: energy minimization method described in section II**; initial**: initial step length for line search algorithm, EXT: quadratic extrapolation from previous step, MIN: minimum between a single Newton step and EXT, MIN1 minimum(EXT,1); **search**: line search algorithm: MT Moré Thuente, MA modified Armijo, SB simple backtracking; **details**: equations or algorithms used.

| method | initial | search | detail |
| --- | --- | --- | --- |
| LBFGS | MIN1 | MT | limited memory Broyden-Fletcher–Goldfarb-Shanno method, "Nocedal and W (1989)" |
| LBFGS | MIN | SB | Same as above, but line search as in algorithm 2 |
| PRP+ | EXT | MA | conjugate gradient, eq. (19), line search as in "Bartholomew-Biggs (2008)" |
| PRP+ | MIN | MA | Same as above, initial step length for line search according to eq. (27) |
| PRP+ | MIN | SB | Same as above but line search according to algorithm 2 |
| TMPRP+ | EXT | MA | conjugate gradient, eq. (20), line search as in "Bartholomew-Biggs (2008)" |
| TMPRP+ | MIN | MA | Same as above, initial step length for line search according to eq. (27) |
| TMPRP+ | MIN | SB | Same as above but line search according to algorithm 2 |
| LMV+ | EXT | MA | conjugate gradient, eq. (21), line search as in "Bartholomew-Biggs (2008)" |
| LMV+ | MIN | MA | Same as above, initial step length for line search according to eq. (27) |
| LMV+ | MIN | SB | Same as above but line search according to algorithm 2 |
| ZA | EXT | MA | conjugate gradient, eq. (22), line search as in "Bartholomew-Biggs (2008)" |
| ZA | MIN | MA | Same as above, initial step length for line search according to eq. (27) |
| ZA | MIN | SB | Same as above but line search according to algorithm 2 |
| TaS | MIN | SB | conjugate gradient, eq. (23), line search according to algorithm 2 |
| HuS | MIN | SB | conjugate gradient, eq. (24), line search according to algorithm 2 |

If we switch off the computation of the magnetostatic field our sample reduces to 9 isolated hard magnetic grains whose switching field is given by the Stoner-Wohlfarth formula "Stoner and Wohlfarth (1948)". Owing to the slight misalignment the grains have different switching fields. This leads to steps in the demagnetization curve. A proper numerical scheme shall find all these steps even if the switching field between two grains differs only by a small amount. When a grain switches, the system accesses a different local minimum. During energy minimization we have to track all the local minima along the demagnetization curve. The solid line of figure 2 gives the demagnetization curve according to the analytic switching field.

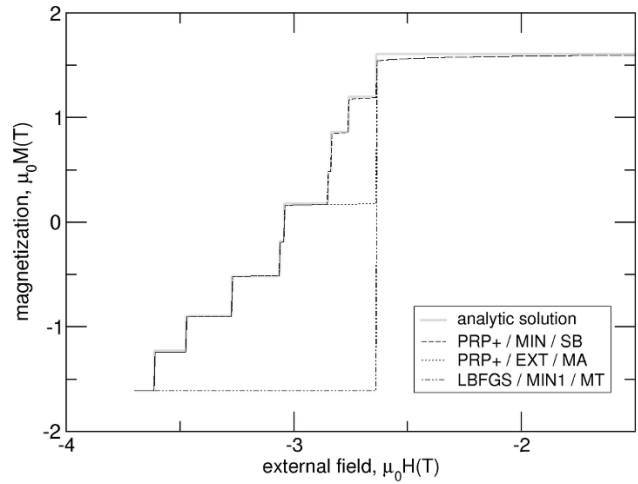

FIG. 2. A model system of 9 isolated grains of a permanent magnet serves a test case for the algorithm. The demagnetization curve can be computed analytically (solid line). Standard method optimization methods such as a limited memory quasi Newton method (LBFGS/MIN1/MT) and a conjugate gradient method (PRP+/EXT/MA) may miss kinks in the demagnetization curve. Conjugate gradient methods with special line search (PRP+/MIN/SB) work well. For the particular numerical models used see Table I. Please note that the analytic solution does not include the slight reversible rotation of the magnetization.

For computing the demagnetization curves in figure 2 we started with an external field $H_{ext} = 0$. Then we decreased the external field in steps of $\mu_0 \Delta H_{ext} = 0.005$ T. The final field value was $\mu_0 H_{ext} = -3.7$ T. The magnetic state from the previous field is used as initial guess for energy minimization (algorithm 1). In figure 2 we compare the analytic solution (solid line) with standard minimization algorithm as discussed in textbooks: A conjugate gradient method "Bartholomew-Biggs (2008)" (dotted line), a limited memory quasi-Newton method "Nocedal and Wright (2006)" (dashed-dotted line). These method miss some kinks in the demagnetization curve. A conjugate gradient method with line search algorithm 2 (dashed line) correctly tracks find all steps in the demagnetization curve.

We tested various combinations of conjugated gradient and line search methods. In table II we summarize the results. Most method that use a back tracking algorithm only can find all kinks in the demagnetization curve. If the line search algorithm also increases the step like in the More Thuente line search or the modified Armijo line search the algorithm may miss the next local minimum and jump to a different local minimum. This leads to too big a step in the demagnetization curve. Depending on where the missed steps occur along the demagnetization curve, missed steps can lead to a wrong coercive field. The 5[th] column in Table II gives the error in the coercive field owing to missed local minima of the energy. This may also happen with the TMPRP+ conjugate gradient method and backtracking line search. We found methods that are fast (low number of function evaluations) and correct (zero kinks missed in the demagnetization curve). The results show that modified conjugate gradient methods that enforce conjugacy between successive search directions work best in terms of the number of function evaluations.

TABLE II. Performance of various energy minimization schemes (see Table I) for the test problem. In order to compare with an analytical solution the magnetostatic field is switched off for the results presented in this table. **misses**: kinks missed in the demagnetization curve. The algorithm is correct if this value is 0. **$\mu_0 \Delta H_c$:** Error in the coercive field owing to missed kinks; **iterations**: average number of iterations per field step, **eval**: average number of function evaluations per field step.

| method | initial | search | misses | $\mu_0 \Delta H_c$(T) | iterations | eval |
|--------|---------|--------|--------|----------------------|------------|------|
| LBFGS  | MIN1    | MT[a]  | 8      | 0.405                | 23         | 15   |
| LBFGS  | MIN     | SB[b]  | 6      | 0.280                | 15         | 30   |
| PRP+   | EXT     | MA     | 3      | 0                    | 30         | 67   |
| PRP+   | MIN     | MA     | 2      | 0                    | 30         | 90   |
| PRP+   | MIN     | SB     | 0      | 0                    | 37         | 75   |
| TMPRP+ | EXT     | MA     | 5      | 0.280                | 28         | 61   |
| TMPRP+ | MIN     | MA     | 3      | 0                    | 30         | 90   |
| TMPRP+ | MIN     | SB     | 1      | 0                    | 31         | 63   |
| LMV+   | EXT     | MA     | 5      | 0                    | 28         | 61   |
| LMV+   | MIN     | MA     | 5      | 0.405                | 28         | 84   |
| **LMV+** | **MIN** | **SB** | **0** | **0**             | **31**     | **63** |
| ZA     | EXT     | MA     | 3      | 0                    | 30         | 65   |
| ZA     | MIN     | MA     | 0      | 0                    | 31         | 94   |
| ZA     | MIN     | SB     | 0      | 0                    | 31         | 64   |
| TaS    | MIN     | SB     | 0      | 0                    | 32         | 66   |
| HuS    | MIN     | SB     | 1      | 0                    | 33         | 68   |

[a] $c_1 = 0.0001$ in (17), $c_2 = 0.9$ in (18)
[b] $c_1 = 0.0001$ in (28)

Limited memory quasi-Newton method are expected to perform better than the conjugate gradient methods "Liu and Nocedal (1989)". Unfortunately, standard schemes that use the Moere-Thuente line search fail to reproduce the analytic demagnetization curve of figure 2. We also tried the limited memory quasi-Newton method with the line search method given by algorithm 2. Here the initial step length is estimated by the local curvature and the step length can only decrease. However, then the curvature condition (24) is often violated and the updates are skipped. The search directions are computed with a bad approximations of the inverse hessian matrix. The algorithm fails to reproduce the demagnetization curve correctly.

Having identified suitable minimization algorithms we compute the demagnetization curve of the permanent magnetic grains under the influence of the self-demagnetizing field. "Grönefeld and Kronmüller (1989)" showed that strong demagnetizing fields occur near the edges and the corners of a grain. These fields drastically reduce the coercive field "Bance et al. (2014)". Figure 3 compares the computed demagnetization curves with magnetostatics switched on and off in the micromagnetic solver. The difference between the two curves has to be attributed to demagnetizing effects and magnetostatic interactions. The maximum misalignment angle of the grains is only 7.5 degrees. The local angle between the demagnetizing field and the anisotropy axes near corners and edges will be much higher. "Thielsch et al. (2013)" showed that magnetization reversal will start at the point where the sum of the external field and the demagnetizing field exceeds the local Stoner-Wohlfarth switching field. All grains but the one which is perfectly aligned (misalignment angle of only 0.1 degrees) switch at the same value of the external field. The demagnetizing field reduces the coercivity by 1.4 T.

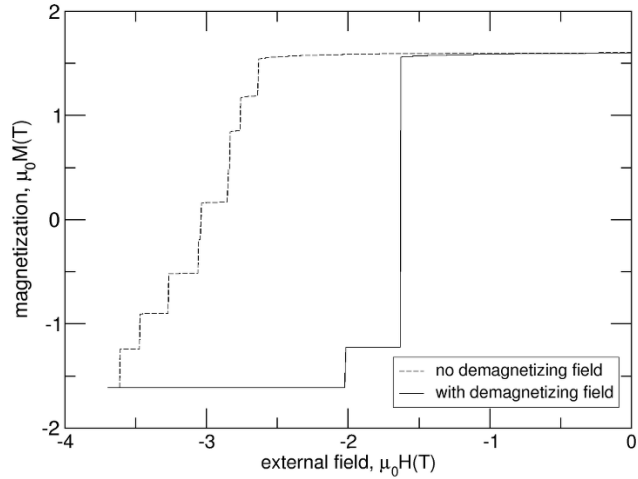

FIG 3. Influence of the demagnetizing field on magnetization reversal in $NdFe_{12}$ based permanent magnets at a temperature of 450 K. For the simulation of the dashed line the demagnetizing field was switched off. The solid line is the result of full micromagnetic simulations including magnetostatics. The granular sample used for the simulations is shown in figure 1.

The performance of the different solvers was also evaluated for full micromagnetic simulations including magnetostatics. In order to avoid very short search directions |**d**| it turned out to be useful to restart the conjugate gradient iterations periodically for the LMV+ method. Table III gives the number of average number of iterations per field step, the average number of function evaluations per field step, and the average number of iterations per field step for solving the linear system for the magnetostatic field. All methods perform. All methods were found to perform equally well. On average there are 23 conjugate gradient iterations per field step, 2 function evaluations per conjugate gradient iteration, and 27 iterations of the linear solver for the magnetic scalar potential per function evaluation.

TABLE III. Performance of various energy minimization schemes (see Table I) for the permanent magnet and the soft magnetic thin film. ; $\Delta H_{ext}$: Field interval between two points in the demagnetization curve; **steps**: total number of field steps. **iterations**: average number of iterations per field step; **eval**: average number of function evaluations per field step; **demag iter**: average number of iterations per field step for solving the linear system for the magnetic scalar potential

| method | initial | search | sample | $\mu_0\Delta H_{ext}$(T) | steps | iterations | eval | demag iter |
|---|---|---|---|---|---|---|---|---|
| PRP+ | MIN | SB | hard magnet | 0.005 | 741 | 23 | 47 | 1293 |
| LMV+[a] | MIN | SB | hard magnet | 0.005 | 741 | 23 | 47 | 1285 |
| ZA | MIN | SB | hard magnet | 0.005 | 741 | 23 | 47 | 1273 |
| TaS | MIN | SB | hard magnet | 0.005 | 741 | 23 | 47 | 1274 |
| LBFGS | MIN1 | MT[b] | soft film | 0.001 | 101 | 228 | 230 | 9156 |
| LMV+[a] | MIN | SB | soft film | 0.001 | 101 | 534 | 1069 | 31019 |
| ZA | MIN | SB | soft film | 0.001 | 101 | 282 | 565 | 16628 |

[a] with restart every 50 steps
[b] $c_1 = 0.0001$ in (17), $c_2 = 0.9$ in (18)

### F. Soft magnetic thin film elements

We also checked the performance of the different solvers for the micromagnetic standard problem #1 "McMichael and Donahue (1997)". The sample is a permalloy film with dimensions 1000×2000×20 nm³. The material parameters are a magnetization of $\mu_0 M_s = 1.01$ T, an exchange constant of $A = 13$ pJ/m. There is uniaxial magneto-crystalline anisotropy along the long axis of the sample with an anisotropy constant of $K = 500$ J/m³. We compute the demagnetization curve with the external field applied at one degree with respect to the long axis of the sample. The external field is varied from +0.05 T to −0.05 T.

For our simulations with use a tetrahedral finite element mesh with a mesh size of 5 nm. We choose a field step of $\mu_0\Delta H_{ext} = 0.001$ T. Figure 4 shows the computed demagnetization curve. We obtain the very same curve for all minimization methods. It seems that for this particular sample the result is less sensitive to the features of line search algorithm. The performance of algorithms in terms of iterations and function evaluations per field step is listed in table III. The limited memory quasi-Newton solver performs best showing the lowest number of iterations. In contrast to the permanent magnet example there is a significant performance difference between the different conjugate gradient methods. The fastest conjugate gradient method is the modified Hestenes-Stiefel method (22). It is about a factor 1.5 times slower than the limited memory quasi Newton method. The conjugate gradient variants PRP+ and TaS require more than $2\times10^5$ conjugate iterations at an external field of $\mu_0 H_{ext} = 0.03$ T and of $\mu_0 H_{ext} = -0.008$ T, respectively. Thus they are not suitable for practical simulations.

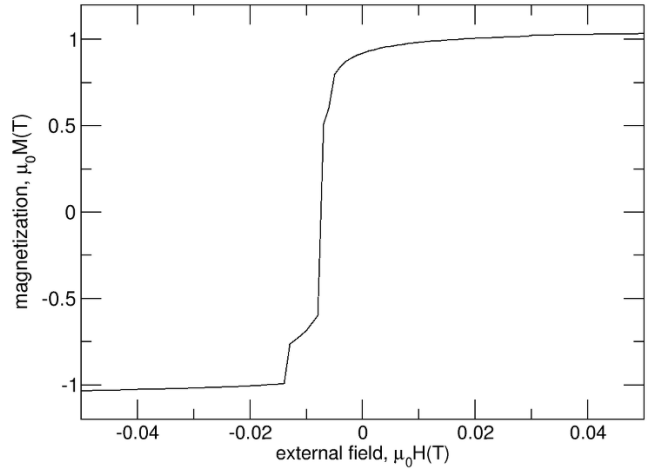

FIG. 4. Computed demagnetization curve for the field applied one degree of the long axes for the micromagnetic standard problem #1. For all investigated solvers (see Table III) the computed results are identical.

**IV. CONCLUSIONS**

We compared various conjugate gradient methods for their use in static micromagnetic simulations. Accurate and efficient computation of the coercive field of hard magnetic grains requires modification of the line search. The results show that a single one dimensional Newton step followed by backtracking is a successful line search procedure in combination with variants of the conjugate gradient method that preserve conjugacy between successive search direction. With these methods the minima along the demagnetization curve for had magnetic materials can be accurately tracked despite the use of an inexact line search algorithm.

Applying the newly developed method to magnetization reversal in $NdFe_{12}$ based permanent magnets show that demagnetizing effects reduce the coercive field in high temperature applications.

**ACKNOWLEDGMENTS**

This work was supported by the Austrian Science Fund (FWF): F4112 SFB ViCoM and the pioneering program "Development of magnetic material technology for high-efficiency motors" (2012–) commissioned by the New Energy and Industrial Technology Development Organization (NEDO).